# Standard estimators cannot represent fault-tolerant workloads at measured error rates: evaluated, evidence-based uncertainty for quantum resource estimation

**Furqan Nasir**[1,2], **Sher Jeel Ahmad**[3]

[1] National University of Computer and Emerging Sciences (FAST-NUCES), Islamabad, Pakistan. [2] City University of Science and Information Technology, Peshawar, Pakistan. [3] University of Engineering and Technology, Peshawar, Pakistan.

Corresponding author: Furqan Nasir (furqannr@gmail.com).

**Abstract.** Estimates of the quantum resources needed to run fault-tolerant algorithms are almost always reported as single numbers, even though they rest on uncertain hardware parameters and on cost models that disagree. We present an open, tool-agnostic framework that propagates evidence-based priors over fault-tolerant hardware parameters through five surface-code cost models, adds correlated-input sensitivity analysis, and evaluates the resulting intervals with a pre-registered three-layer protocol. Applied to RSA-2048, ECC-256, AES-256, and a Hubbard-model simulation on a superconducting surface-code architecture, it produces three coupled findings that together undermine the single-number convention. Propagating the error rates hardware has actually demonstrated, whose median across five large-array devices is about $4 \times 10^{-3}$ rather than the conventional $10^{-3}$, widens the 90% physical-qubit interval to roughly forty times and lifts its median about 4.6 times above the conventional point estimate. Against the same inputs, five independently structured cost models, including two third-party estimators (the Azure Quantum Resource Estimator and Qualtran) that agree with each other to within about ten percent, disagree by a stable factor of two, a structural uncertainty that no single tool reveals. Most consequentially, both third-party estimators exceed their representable envelope for 80 to 100 percent of the evidence-based parameter space, and the AES-256 gate count overflows one tool's counter outright, so these workloads cannot be represented at measured error rates at all. The optimistic convention conceals every one of these effects. The uncertainty machinery we use is standard. The contribution is its evidence-based application, the pre-registered evaluation, and the finding that standard tools break down precisely where hardware operates. We release the framework, the prior library, and the analysis code.

# 1. Introduction

Estimates of the quantum resources needed to run Shor's algorithm against RSA-2048 or to simulate a strongly correlated material are load-bearing inputs to real decisions. Standards bodies and governments use them to time the migration to post-quantum cryptography (Joseph et al. 2022), because under the harvest-now-decrypt-later threat, data harvested today can be decrypted once a cryptographically relevant quantum computer exists, so the migration deadline depends on when that machine is expected. Hardware roadmaps and claims of quantum advantage cite the same estimates. Yet a resource estimate is almost always reported as a single set of numbers, so many logical qubits, so many Toffoli gates, and so many hours. Yet it rests on hardware and architecture parameters that are uncertain, and on modelling choices that differ from one group to the next. A point estimate cannot express how much of its value is a genuine prediction and how much is a fixed assumption.

The most consequential of these parameters is the physical two-qubit gate error rate, because the surface-code overhead grows sharply as the error rate approaches the code threshold. The community's planning convention fixes this rate at $10^{-3}$. The large-array superconducting devices that have actually been characterised tell a different story. Across five such devices, namely Google Willow, IBM Heron, USTC Zuchongzhi-3, and Google Sycamore, the median average two-qubit gate error is about $4 \times 10^{-3}$, four times the planning value. No published resource estimate propagates the distribution of measured error rates into the resource number, so the gap between the optimistic convention and demonstrated hardware never appears in the estimates that decisions rely on. Our central empirical result is what happens when that gap is propagated honestly.

The tools for computing these estimates have grown steadily more realistic. Analytic surface-code scaling (Fowler et al. 2012), lattice-surgery tile accounting (Litinski 2019), the Azure Quantum Resource Estimator (Beverland et al. 2022), and recent frameworks such as SPARO (Kan et al. 2025), FLASQ (Huggins et al. 2025), and Qualtran (Harrigan et al. 2024) each refine how logical operations map to physical cost. Every one of them, however sophisticated, still reports a point value or a handful of fixed scenarios. None propagates a distribution over the uncertain inputs, and none checks whether its output interval is calibrated against later evidence. That is the gap this paper closes.

We built an open, tool-agnostic uncertainty-quantification layer that propagates evidence-based priors over fault-tolerant hardware parameters through several resource estimators, adds correlated-input sensitivity analysis, and evaluates the resulting intervals with a pre-registered protocol. Applied to canonical workloads on a superconducting surface-code architecture, it produces three coupled findings.

**Contributions.**

1. **An evaluated-uncertainty framework.** An open, tool-agnostic layer that propagates a documented, evidence-based prior library through five surface-code cost models, with correlated-input Shapley sensitivity and a pre-registered three-layer evaluation registered before the confirmatory runs (Tables 1 and 2, Section 3).

2. **Parametric uncertainty is large.** At measured superconducting error rates the 90% resource interval spans about forty times, and its median sits roughly 4.6 times above the point estimate computed at the $10^{-3}$ planning convention (Figure 1, Table 1).
3. **Structural disagreement is real and tool-corroborated.** Five independently structured cost models, including two third-party estimators that agree with each other to within about ten percent, bracket a consistent factor-of-two spread at matched inputs (Figure 2, Table 2).
4. **Standard estimators cannot represent the measured-error regime.** Two independently built third-party tools exceed their representable envelope for 80 to 100 percent of the evidence-based parameter space, a fact the $10^{-3}$ convention hides completely (Figure 3, Table 2).

We are explicit about what is new here and what is not. The uncertainty-propagation machinery we use, namely Monte Carlo sampling with global Shapley and Sobol sensitivity analysis, is standard and, taken on its own, textbook. The contribution is not a new statistical method. It is the application of evidence-based uncertainty quantification to fault-tolerant resource estimation, the construction of a documented prior library from primary hardware measurements, and the pre-registered evaluation of the resulting intervals. Above all, it is the empirical finding that standard estimators break down at measured error rates. Stating this plainly matters, because it locates the contribution in the domain and the evidence rather than in the mathematics, and it separates this work from two adjacent but distinct lines of research. One line quantifies uncertainty in the outputs of near-term noisy devices, which concerns statistical error on a running machine rather than the planning uncertainty of a future one. Another forecasts quantum-computing timelines from historical hardware trends, which projects capability forward in time rather than propagating parameter uncertainty through a fixed cost model at a fixed algorithm. Our question is narrower and, for a decision-maker sizing a threat, more direct. Given what hardware has actually demonstrated, how uncertain is the resource cost of a fixed algorithm, and can standard tools even represent it.

We report results for the superconducting surface-code architecture. This is a deliberate scope choice rather than a limitation of the framework. Superconducting hardware is at present the only platform with enough independent large-array error-rate measurements to support an evidence-based prior for the dominant parameter, and building priors from thin or non-representative evidence would undercut the paper's own argument. The framework is platform-tagged and the prior library already carries neutral-atom and trapped-ion entries, so extension is a matter of evidence rather than of new machinery. The rest of the paper is organised as follows. Section 2 relates the work to existing estimators and to uncertainty-quantification methodology. Section 3 describes the framework, the evidence-based prior library, the correlated-input sensitivity analysis, and the pre-registered evaluation design. Section 4 reports the three findings and evaluates them. Section 5 discusses their implications for cryptographic-threat planning and for advantage claims, states the limitations, a

## 2. Related work

Our work sits at the intersection of three lines of research: fault-tolerant resource estimation, uncertainty quantification and sensitivity analysis, and the small but growing literature on uncertainty in quantum-computing forecasts.

**Resource estimators and cost models.** A sequence of increasingly realistic estimators converts logical algorithm counts into physical resources for a surface-code architecture. Fowler et al. (2012) gave the textbook scaling that underlies most analytic estimates. Litinski (2019) introduced the lattice-surgery tile accounting that we use as a structurally distinct cost model. Beverland et al. (2022) developed the model behind the Azure Quantum Resource Estimator, which we run as a validated third-party tool. For the specific workloads we study, Gidney and Ekerå (2021) gave the exact abstract-circuit counts for factoring RSA-2048, Roetteler et al. (2017) did the same for elliptic-curve discrete logarithms, and Jaques et al. (2020) for Grover search on AES. Gidney (2025) has since reduced the RSA-2048 estimate by more than an order of magnitude through algorithmic and architectural change. More recent tools such as SPARO (Kan et al. 2025) and FLASQ (Huggins et al. 2025) refine the spacetime accounting further, and the Qualtran library (Harrigan et al. 2024) provides a second independent implementation of the Gidney and Fowler magic-state cost model, which we use as our fifth model. Closest in spirit to our study, Gheorghiu et al. (2024) report resource estimates for cryptanalysis across a range of physical error rates, though they produce scenario snapshots rather than a propagated, evidence-based distribution that is then evaluated. Every one of these estimators, however sophisticated, reports a point value or a small number of fixed scenarios. None propagates a distribution over the input parameters, and none evaluates the calibration of its output. The present work adds exactly that layer on top of them rather than proposing a replacement.

**Uncertainty quantification and sensitivity analysis.** The propagation and attribution of input uncertainty is a mature field outside quantum computing. Variance-based global sensitivity analysis through Sobol indices (Saltelli et al. 2008) is the standard tool for identifying which inputs drive an output, but Sobol indices lose their clean interpretation when inputs are correlated. For that case we use Shapley effects, which fairly allocate shared variance among correlated inputs (Owen 2014; Iooss and Prieur 2019). Constructing priors from expert and measured evidence draws on the established practice of structured expert judgement and elicitation (O'Hagan 2019). What is new is not these methods but their application to fault-tolerant resource estimation. Here the inputs are hardware parameters with a real and growing measurement record, so the output interval can be checked against later evidence rather than merely reported.

**Uncertainty in quantum-computing forecasts.** Two adjacent efforts share our concern with uncertainty but answer different questions. One body of work quantifies uncertainty in the outputs of near-term noisy devices (Bennink et al. 2026), which is a statement about statistical and systematic error on a machine that is running now, not about the planning uncertainty of a machine that does not yet exist. A separate effort forecasts quantum-computing capability from historical hardware trends (Sevilla and Riedel 2020), projecting when a given task will become feasible. Our question is orthogonal to both. We hold the algorithm and the architecture fixed, propagate the measured uncertainty in the hardware parameters, and ask how wide and how well-calibrated the resulting resource interval is, and whether standard too

# 3. Framework and methods

We treat resource estimation as an uncertainty-propagation problem. A resource estimator maps a set of hardware and architecture parameters to a resource cost such as physical-qubit count and runtime. The parameters are not known exactly. They are estimated from a finite and sometimes discordant body of experimental evidence, and different published cost models convert the same parameters into costs in different ways. Our framework quantifies both of these effects. It propagates evidence-based parameter distributions through several estimators, records where an estimator cannot represent a given input, and evaluates the resulting intervals against a pre-registered protocol.

We can state this formally. A single draw of hardware and architecture parameters is a vector

$$\mathbf{x} = [\, p_{\text{phys}},\ t_{\text{cycle}},\ \tau_{\text{factory}},\ \gamma_{\text{route}},\ t_{\text{decode}} \,],$$

which collects the physical two-qubit gate error rate, the surface-code cycle time, the magic-state factory throughput, the routing overhead, and the decoder reaction time. For a fixed workload $W$, a resource estimator is a map

$$\mathbf{y} = f(\mathbf{x}, W),$$

whose output $\mathbf{y}$ holds the physical-qubit count, the code distance, the runtime, and the factory requirement. Each parameter carries an evidence-based prior $p(\mathbf{x})$, so the resource cost is not a single point but a distribution,

$$p(\mathbf{y}) = \int f(\mathbf{x}, W)\, p(\mathbf{x})\, \mathrm{d}\mathbf{x}.$$

The framework estimates this distribution by pushing the prior through each of the five cost models in turn,

$$p(\mathbf{x}) \ \rightarrow\ \{f_m\}_{m=1}^{5} \ \rightarrow\ p(\mathbf{y}),$$

and reads three quantities off the result: the spread of $p(\mathbf{y})$ for a single model, the disagreement among the $f_m$ at matched inputs, and the fraction of draws that each $f_m$ cannot represent. What we report is therefore a prediction interval over the resource distribution, not a confidence interval on a fitted parameter.

## 3.1 A tool-agnostic uncertainty-quantification layer

The layer is organised around a single input schema and a single output schema, so that one uncertainty analysis can be run across many back-end estimators without re-encoding the problem for each one. A draw of the hardware parameters is represented by a fixed record, with one named field per parameter and a platform tag. Each back-end is a thin adapter that translates this record into the tool's native inputs and translates the tool's output back into a common resource record. Adapters that cannot compute a quantity report it as missing rather than substituting a zero, so that gaps stay visible in the output distribution.

Parameter draws are generated by inverse-transform sampling from the prior library. Three samplers are available, namely plain Monte Carlo, Latin-hypercube sampling, and a Sobol quasi-random sequence. Truncation is applied by restricting the sampled quantile range and mapping it back through the inverse cumulative distribution, which preserves the space-filling structure of the Latin-hypercube and Sobol sequences rather than discarding out-of-range points. A single seed determines a run completely, and long sweeps are split into independent shards so that a run interrupted by a session time limit resumes from the first missing shard. Draws for which a back-end fails are retained as missing-valued rows with a success flag set to false. This turns the failure region into observable data. We use it directly in Section 4 to measure how often a standard estimator cannot represent a workload at measured error rates.

## 3.2 Five cost models grounded in published constructions

The framework runs five surface-code cost models on identical parameter draws. Three are transparent analytic or tile-counting models that we implemented from published constructions, and two are third-party estimators developed by other groups. The purpose of using more than one model is to measure structural, or model-form, uncertainty. This is the disagreement that remains when the same inputs are converted to costs by different but individually defensible accounting schemes.

On a common set of representable draws we summarise this disagreement as the dispersion of the five model outputs at matched inputs,

$$U_{\text{model}} = \text{Var}(f_1, f_2, \dots, f_5).$$

Section 4 reports it as the ratio of the largest to the smallest model median, so that it reads on the same multiplicative scale as the resource itself rather than as a raw variance.

To make this measurement meaningful rather than an artefact of arbitrary constants, every model constant is taken from a specific published source rather than tuned by us.

The first model is a textbook analytic surface-code estimator following the scaling of Fowler et al. (2012) and Beverland et al. (2022). It sets the code distance from the physical error rate and a per-round logical-error target, allocates two physical qubits per data qubit per code cell through the standard $2d^2$ patch rule, and sizes magic-state factories to feed the Toffoli stream. The second model follows the lattice-surgery tile accounting of Litinski (2019). It uses fast data blocks of two tiles per logical qubit and roughly fifteen-tile Toffoli or T factories, which is a genuinely different spatial layout from the first model. The third model follows the compact-block accounting associated with Fowler and Gidney, using denser data blocks of about one and a half tiles per logical qubit and a larger factory footprint. The three transparent models share only the physics that they must share, namely the exponential suppression of the logical error rate with code distance. They differ in how data qubits and factories are laid out, which is exactly the axis along which published schemes disagree.

The fourth model is the Azure Quantum Resource Estimator (Beverland et al. 2022), which we run locally through its Q# package. It is a validated third-party tool, but it caps the code distance at fifty by default. Draws whose required distance exceeds that cap raise an estimator error and are recorded as censored. The fifth model is the Gidney and Fowler catalysed-CCZ-to-2T cost model as implemented in Qualtran (Gidney and Fowler 2019). It is a second, independently written third-party estimator. It uses a fixed distillation factory with level-one and level-two code distances of fifteen and thirty-one and selects the data-block distance from the remaining error budget. When the fixed factory's total distillation error exceeds the budget, the draw is infeasible and is recorded as censored. The two third-party tools therefore have different representable envelopes set by different mechanisms, which is useful. Agreement between them is not built in, and where it occurs it is evidence rather than assumption.

We anchor the calibrated tool to the literature at the conventional planning point. At a physical error rate of $10^{-3}$ and a one-microsecond cycle, the Azure estimator returns about $2.19 \times 10^7$ physical qubits for RSA-2048 on the exact Gidney and Ekerå counts, which is within a factor of 1.1 of the published estimate of roughly $2 \times 10^7$ (Gidney and Ekerå 2021). On matched inputs across the representable range, the three transparent models track the Azure estimator by stable factors, so extrapolating them into the censored high-error region is supported by the overlap where both can be computed.

## 3.3 An evidence-based prior library

The prior for each parameter is fitted from a documented table of published measurements and modelling values rather than assumed. We report results for the superconducting two-dimensional nearest-neighbour surface-code platform, which is the only platform with enough independent large-array error-rate measurements to support evidence-based priors at present. Each parameter is represented by a lognormal distribution in its canonical unit, fitted by weighted maximum likelihood and truncated to a physical support where appropriate. Table 2 lists the fitted priors.

Each parameter $x_i$ is modelled by a truncated lognormal prior,

$$x_i \sim \mathrm{LogNormal}(\mu_i,\, \sigma_i^2),$$

with $\mu_i$ and $\sigma_i$ set by weighted maximum likelihood on the evidence table. Treating the parameters as conditionally independent in the base propagation, the joint prior factorises as

$$p(\mathbf{x}) = \prod_{i=1}^{d} p\,(x_i), \qquad d\, = 5,$$

and the one modelled dependence, between the error rate and the cycle time, enters only through the Gaussian copula used for the sensitivity analysis in Section 3.4.

The dominant parameter is the physical error rate, for which we define the estimand precisely as the average two-qubit gate error of a scalable two-dimensional surface-code array. Six independent sources inform it. Five are large-array surface-code devices, namely Google Willow at $3.0 \times 10^{-3}$, IBM Heron at $2.96 \times 10^{-3}$, USTC Zuchongzhi-3 at $3.8 \times 10^{-3}$, and Google Sycamore at $5.0 \times 10^{-3}$ as an average-gate figure. The sixth and a down-weighted small-scale record from a fluxonium device at $8 \times 10^{-4}$ and the optimistic planning convention of $1 \times 10^{-3}$ complete the set. Because a single value like the Sycamore figure can be quoted either as a Pauli error or as an average gate error, we convert all sources to the average-gate estimand before fitting and record the convention for each. The fitted error-rate prior has a median of $4.0 \times 10^{-3}$ with a log-scale standard deviation of about 0.49. This median sits four times above the $1 \times 10^{-3}$ planning convention, which is the central quantitative tension that the rest of the paper propagates.

The elicitation protocol addresses three well-known hazards. Sources that are not statistically independent, such as several reports from one device generation, are grouped so that they are not counted as separate evidence. Small-scale or single-pair records are down-weighted for representativeness relative to large-array devices. A survivorship factor of 1.3 shifts the error-rate prior worse than the published record, on the reasoning that a typical deployed device underperforms the best reported chip. This factor is a modelling judgment, and rather than fix it we sweep it and report the sensitivity of the conclusions to it. The remaining parameters are informed by fewer sources and are deliberately given wide priors. The cycle time has a median of about $1.08 \times 10^{3}$ nanoseconds, the factory throughput about $2.24 \times 10^{4}$ states per second, the routing overhead about 1.69, and the decoder reaction time about $5.6 \times 10^{3}$ nanoseconds. As Section 4 shows, the width of these three-to-five-source priors has almost no effect on the output interval, because the error rate dominates.

## 3.4 Sensitivity analysis for correlated inputs

To identify which parameters drive the output interval, we compute global sensitivity indices. Because some inputs may be correlated, and because Sobol indices lose their clean variance interpretation under dependence, we take Shapley effects as the primary measure. Shapley effects fairly allocate shared output variance among correlated inputs. We compute the exact five-input Shapley values from all thirty-two subset costs under a Gaussian-copula joint prior, with a correlation parameter coupling the error rate and the cycle time. Sobol indices under independence are reported as the special case. Reporting both lets us state, with a computed basis rather than an assumption, whether correlation materially changes the attribution.

## 3.5 A pre-registered three-layer evaluation

An interval is only useful if it is evaluated rather than merely asserted. We registered the evaluation design publicly before running the confirmatory analyses, at the Open Science Framework under identifier 3d9m2, dated 19 July 2026. The registration fixes the metrics, the estimand definition, the source-weighting rules, and the labelling of each layer as confirmatory or exploratory. This commitment is what lets a reader distinguish a genuine test from a post-hoc description.

The design has three layers. Layer 1 tests numerical propagation correctness. It pushes analytic benchmark distributions with closed-form outputs through every sampler and checks that the recovered quantiles match the closed forms within a pre-registered tolerance. Layer 2 tests prior-predictive coverage by leave-one-source-out. For each held-out measured source, it asks whether that source falls inside the interval predicted by a prior fitted from the remaining sources. Because coverage should be judged by how much each parameter actually drives the output rather than by a naive average across parameters, Layer 2 is reported three ways, namely pooled across all parameters, restricted to the dominant error-rate estimand, and weighted by each parameter's output sensitivity. The sensitivity-weighted metric is the pre-registered primary. Layer 3 is a retrospective hindcast and is labelled exploratory in the registration, because no fault-tolerant machine has executed these workloads and realised ground truth is therefore unobtainable. It uses an expanding window over a dated time series of error-rate measurements to ask whether a prior fitted from earlier evidence would have contained the values reported later, and it decomposes any miss into calibration error versus genuine technological drift.

### 3.6 Workloads and magic-state accounting

We apply the framework to four workloads that fix the algorithm-level inputs. The uncertainty analysis varies hardware and architecture parameters for a fixed algorithm. It does not model algorithmic change. Table 1 lists the workloads with the exact logical-qubit and magic-state counts taken from their published constructions.

| Workload | Logical qubits | Magic-state count | Type | Source |
|---|---|---|---|---|
| RSA-2048 (Shor) | 6189 | $2.624 \times 10^{9}$ | Toffoli | Gidney and Ekerå 2019, 2021 |
| ECC-256 (Shor) | 2330 | $1.26 \times 10^{11}$ | Toffoli | Roetteler et al. 2017 |
| AES-256 (Grover) | 6913 | $6.066 \times 10^{43}$ | T gate | Jaques et al. 2020 |
| Hubbard-model QPE | 1000 | $1.0 \times 10^{10}$ | Toffoli | order-of-magnitude proxy |

Three of the workloads are quoted in Toffoli or CCZ magic states and one, AES-256 under Grover search, is quoted in T gates. The AES-256 count is the full-Grover T-gate figure of $1.36 \times 2^{145}$ from Jaques et al. (2020) with three plaintext-ciphertext pairs. Where a model or tool consumes T states, we use the standard conversion of four T gates per Toffoli in its factory accounting, and we state the convention with each result so that the two magic-state currencies are never silently mixed. The Hubbard-model workload is included as an order-of-magnitude simulation proxy and is labelled as such, since its counts are not pinned to a single exact construction in the way the cryptographic workloads are.

## 4. Results

We report three coupled findings for the superconducting surface-code platform, then evaluate them. Parametric uncertainty is large. Structural disagreement among cost models is real and corroborated by two independent third-party tools. Standard estimators cannot represent the canonical workloads at measured error rates. A pre-registered protocol then tests these intervals, confirming the framework where it is testable and marking honestly where it is not.

## 4.1 Propagating measured error rates yields wide intervals

To quantify how much the single-number convention omits, we first propagate the full evidence-based priors and examine the resulting distribution of physical-qubit counts. At these priors, the resource estimate is a broad distribution rather than a point. Figure 1 shows the RSA-2048 physical-qubit distribution for the three transparent cost models. Table 1 lists the quantiles for all four workloads from two million Monte Carlo draws.

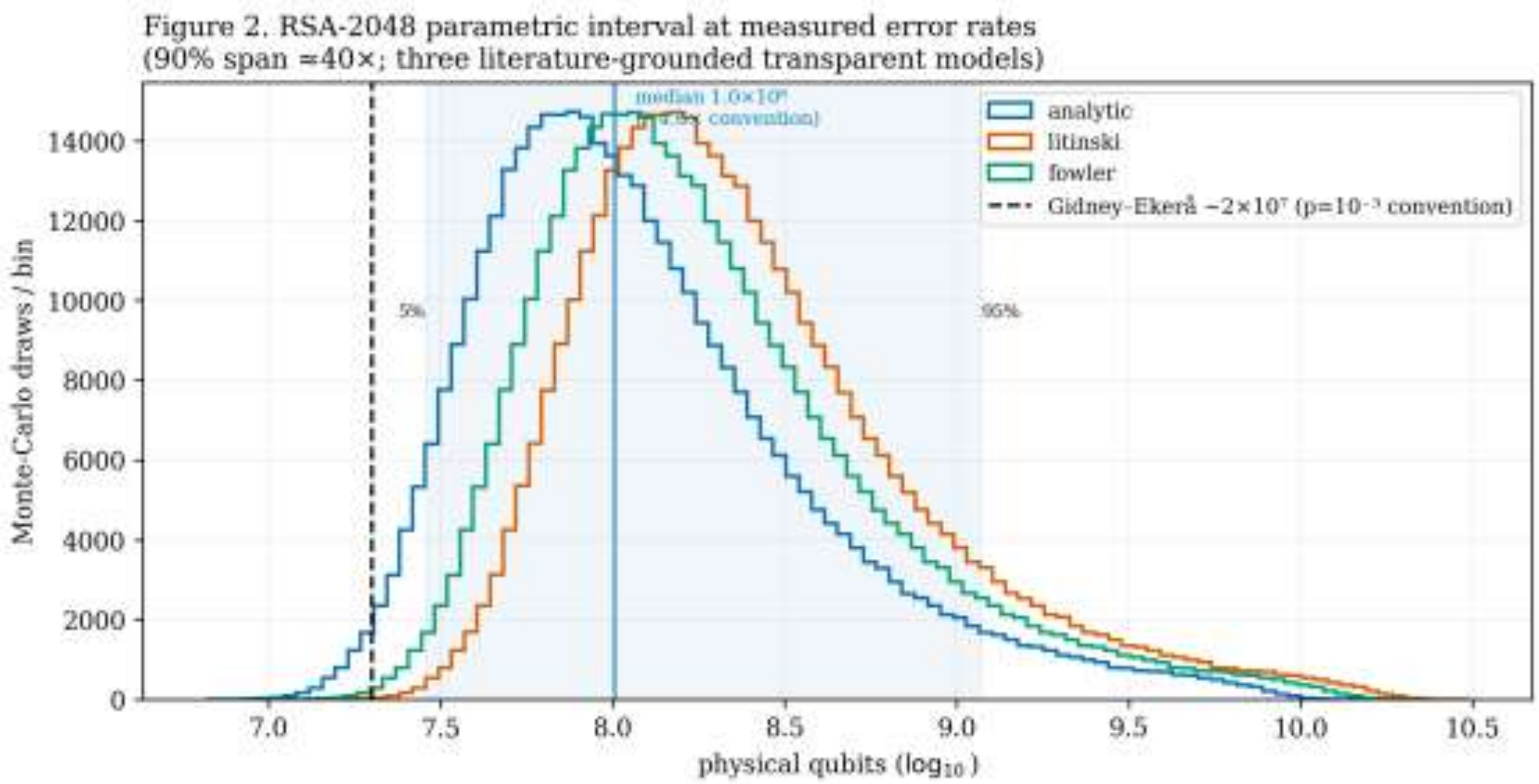

*Figure 1. RSA-2048 physical-qubit distributions under the evidence-based priors for the three transparent models. The 90% interval spans about forty times, and the median near $1.0 \times 10^8$ sits roughly 4.6 times above the Gidney and Ekerå point of about $2 \times 10^7$ computed at the conventional $10^{-3}$ planning error rate (dashed).*

The 90% interval spans about forty times on every workload, measured as the ratio of the 95th to the 5th percentile. For RSA-2048 it runs from $2.86 \times 10^7$ to $1.18 \times 10^9$ physical qubits, a factor of 41. The median of $1.01 \times 10^8$ sits about 4.6 times above the roughly $2 \times 10^7$ point estimate that the community computes at the optimistic $10^{-3}$ planning convention. The same pattern holds for ECC-256 and the Hubbard simulation. AES-256 under Grover search is the extreme case. Its median physical-qubit count is $1.32 \times 10^9$, and our analytic model places its runtime on the order of $10^{32}$ years, which confirms it as the canonical impractical workload. This runtime is a model estimate rather than a measured or previously published value, and we label it as such.

| Workload | 5% | 25% | Median | 75% | 95% | 90% span |
|---|---|---|---|---|---|---|
| RSA-2048 | $2.86 \times 10^7$ | $5.57 \times 10^7$ | $1.01 \times 10^8$ | $2.23 \times 10^8$ | $1.18 \times 10^9$ | 41× |
| ECC-256 | $1.28 \times 10^7$ | $2.50 \times 10^7$ | $4.54 \times 10^7$ | $9.99 \times 10^7$ | $5.28 \times 10^8$ | 41× |
| AES-256 | $3.73 \times 10^8$ | $7.28 \times 10^8$ | $1.32 \times 10^9$ | $2.92 \times 10^9$ | $1.54 \times 10^{10}$ | 41× |
| Hubbard QPE | $4.57 \times 10^6$ | $8.88 \times 10^6$ | $1.61 \times 10^7$ | $3.54 \times 10^7$ | $1.87 \times 10^8$ | 41× |

Table 1. Physical-qubit interval quantiles (two million analytic draws). The 90% span is the ratio of the 95th to the 5th percentile.

The width and the upward shift both come from the same source. The evidence-based error-rate prior has a median of $4.0 \times 10^{-3}$, four times the $10^{-3}$ planning value, and the required code distance grows as the error rate approaches the surface-code threshold. A single point computed at $10^{-3}$ therefore understates both the central estimate and its uncertainty.

## 4.2 Structural disagreement is real and tool-corroborated

Different published cost models convert identical inputs into different costs. To measure this structural uncertainty rather than assert it, we compare all five models on the same draws, restricted to the region that both third-party tools can represent so that the comparison is like for like. Figure 2 and Table 2 report the result.

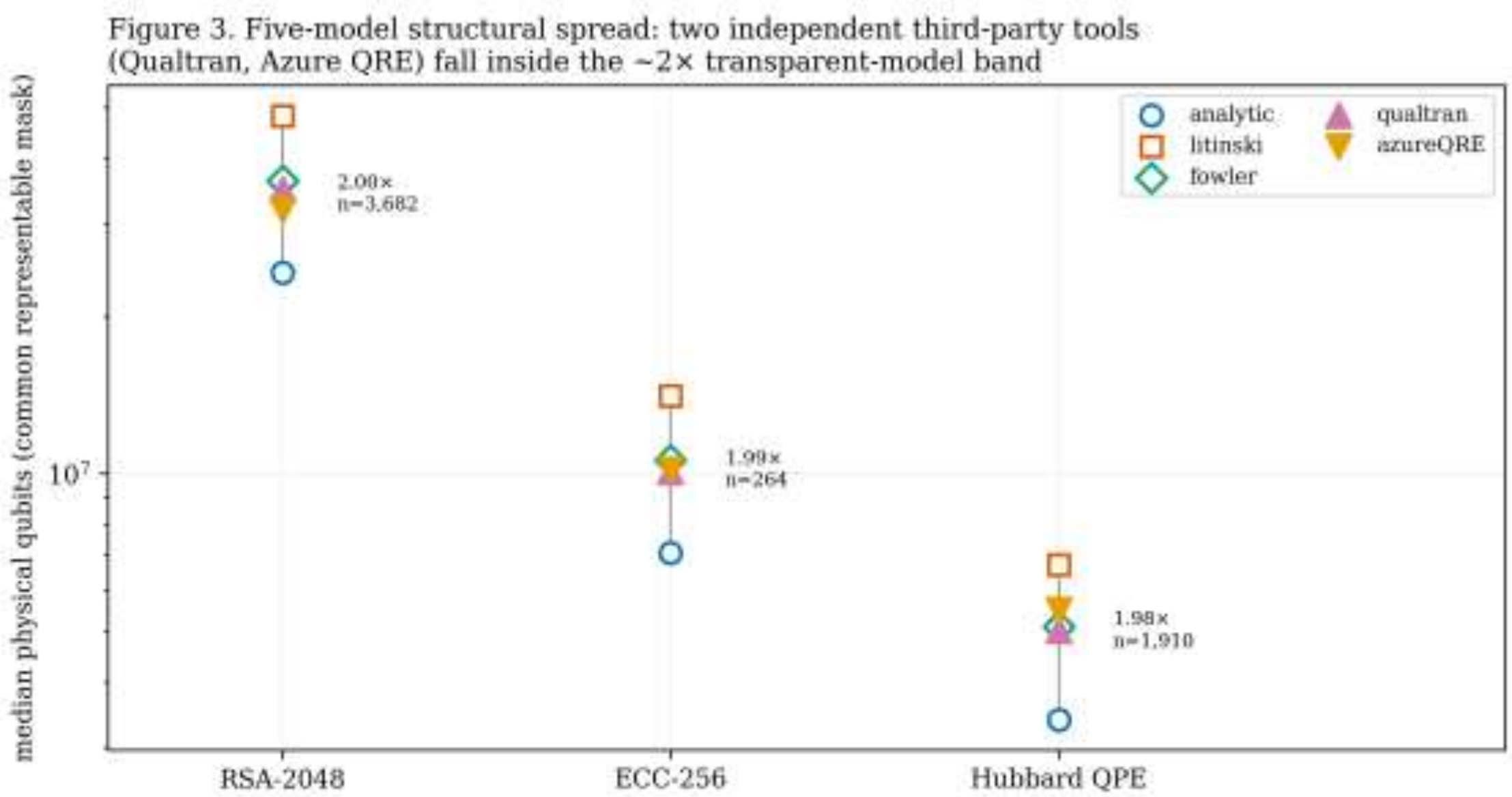

*Figure 2. Median physical qubits for all five models on the common representable mask. The two independent third-party tools (Qualtran, Azure QRE) fall inside the roughly twofold band set by the transparent models and agree with each other to within about ten percent. AES-256 is omitted because both tools censor all of its draws.*

On the common representable mask the five-model spread, defined as the ratio of the largest to the smallest median, is 2.00 for RSA-2048, 1.99 for ECC-256, and 1.98 for the Hubbard workload. The decisive point is the behaviour of the two third-party tools. Qualtran and the Azure estimator were written by different groups and use different code-distance selection and factory accounting, yet they agree with each other to within about ten percent, and to under one percent on ECC-256. For RSA-2048 they return $3.49 \times 10^7$ and $3.19 \times 10^7$ physical qubits. For the Hubbard workload they return $5.03 \times 10^6$ and $5.49 \times 10^6$. Both tools fall inside the band set by the transparent analytic and Litinski models. A factor of two that is bracketed by two independent third-party estimators is a genuine structural-uncertainty band rather than a spread we tuned.

| Workload | analytic | Litinski | Fowler | Qualtran | Azure QRE | spread | n |
|---|---|---|---|---|---|---|---|
| RSA-2048 | $2.42 \times 10^7$ | $4.83 \times 10^7$ | $3.63 \times 10^7$ | $3.49 \times 10^7$ | $3.19 \times 10^7$ | 2.00× | 3,682 |
| ECC-256 | $7.07 \times 10^6$ | $1.41 \times 10^7$ | $1.06 \times 10^7$ | $1.01 \times 10^7$ | $1.02 \times 10^7$ | 1.99× | 264 |
| Hubbard QPE | $3.39 \times 10^6$ | $6.71 \times 10^6$ | $5.11 \times 10^6$ | $5.03 \times 10^6$ | $5.49 \times 10^6$ | 1.98× | 1,910 |

Table 2. Median physical qubits for all five models on the common representable mask, with the max-to-min spread and the number of commonly representable draws. AES-256 is absent because both third-party tools censor all of its draws.

We state the limits of this comparison plainly. Because both tools censor most of the evidence-based space, the common representable mask is small, and it is smallest for ECC-256, where only 264 of the eighty thousand draws are representable by both tools. The concordance is therefore established in the low-error region where a third-party cross-check is possible at all. Within that region the two tools agree, and even the thin ECC-256 comparison agrees to under one percent.

## 4.3 Standard estimators censor the measured-error regime

To measure how often a standard tool simply cannot return an estimate at the measured error rates, we record, from the same run that produced the concordance, the fraction of draws each tool must censor. Figure 3 and Table 2 report this. At the measured error rates, both third-party tools exceed their representable envelope for the large majority of draws, and Qualtran censors even more than the Azure estimator.

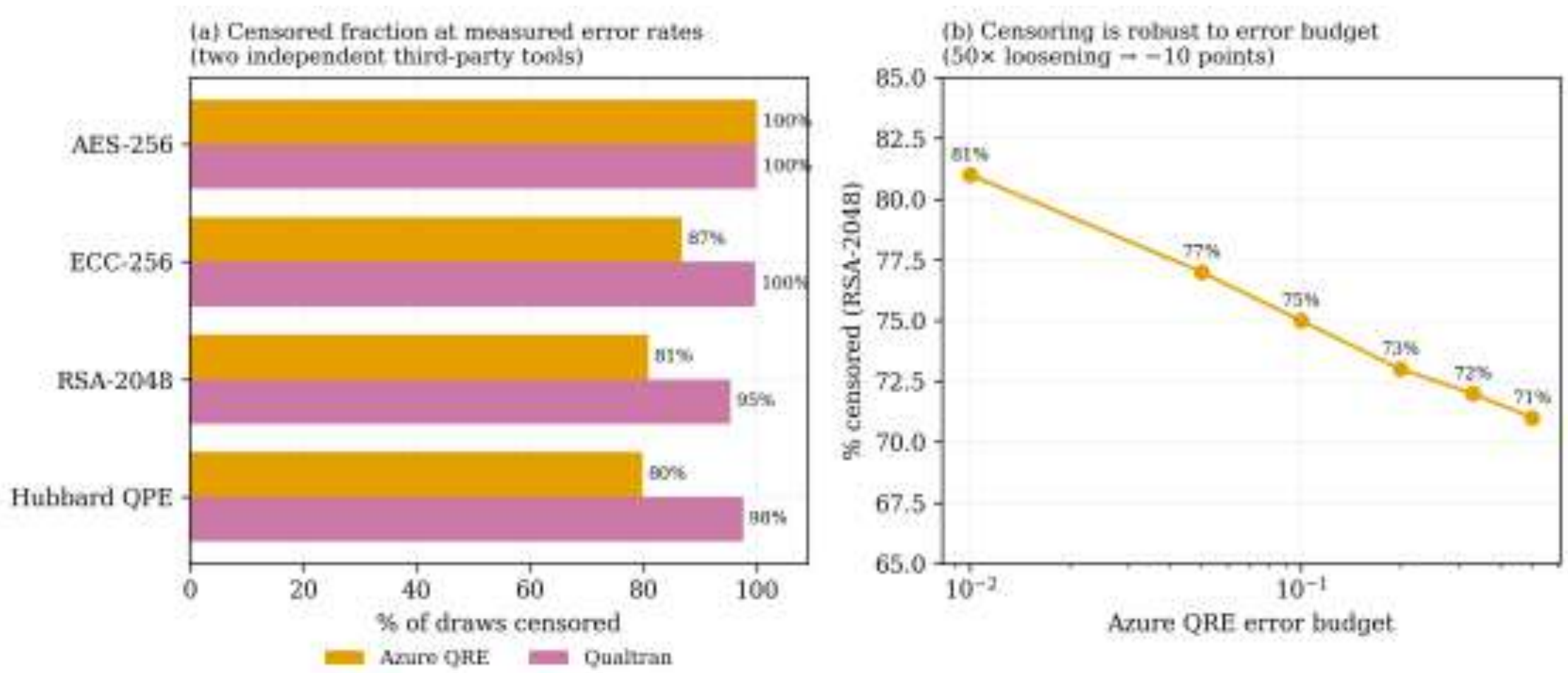

*Figure 3. (a) Fraction of evidence-based draws that exceed each tool's representable envelope, for two independently built third-party tools. (b) For RSA-2048, the Azure censored fraction falls only from 81 to 71 percent as the error budget is loosened fifty times.*

For RSA-2048 the Azure estimator censors 80.9 percent of draws (95 percent Wilson interval 80.6 to 81.2) and Qualtran censors 95.4 percent (95.3 to 95.5). For ECC-256 the figures are 86.8 and 99.7 percent, and for the Hubbard workload 79.8 and 97.6 percent. AES-256 is fully censored by both. With eighty thousand draws per workload, every censored fraction is precise to within 0.3 percentage points at 95 percent confidence, so these are stable estimates rather than sampling artefacts. The effect is not a tight-budget artefact. Loosening the Azure error budget by a factor of fifty, from one percent to fifty percent, lowers the RSA-2048 censored fraction only from 81 to 71 percent.

The two tools reach the wall by different routes, which is why the finding is robust rather than tool-specific. The Azure estimator caps the code distance at fifty. Qualtran instead uses a fixed distillation factory whose total error exceeds the budget once the physical error rate rises above roughly $1.7 \times 10^{-3}$. Qualtran remains feasible only up to about $1.73 \times 10^{-3}$ for RSA-2048, $1.05 \times 10^{-3}$ for ECC-256, and $1.50 \times 10^{-3}$ for the Hubbard workload. The evidence-based error-rate prior has a median of $4.0 \times 10^{-3}$, which sits above both tools' feasibility ceilings. The measured error rates therefore lie outside the region where either standard tool can compute an estimate, for most of the evidence-based parameter space. AES-256 fails in an even more basic way. Its T-gate count of $6.066 \times 10^{43}$ overflows the Azure estimator's sixty-four-bit gate counter, whose maximum is about $1.8 \times 10^{19}$, so the workload cannot even be represented numerically.

We are careful with wording. This is a statement about the default representable envelope of two specific tools, not a claim of physical impossibility. Raising a tool's distance cap or changing its factory would move the boundary. The point is that at the error rates hardware actually demonstrates, and with the tools as their authors ship them, canonical workloads fall outside the representable region, and a point estimate computed at the $10^{-3}$ convention hides this completely.

### 4.4 The pre-registered evaluation

To establish that the intervals are evaluated rather than merely asserted, we apply the pre-registered three-layer protocol, whose metrics and labels were fixed before these runs (Figure 4).

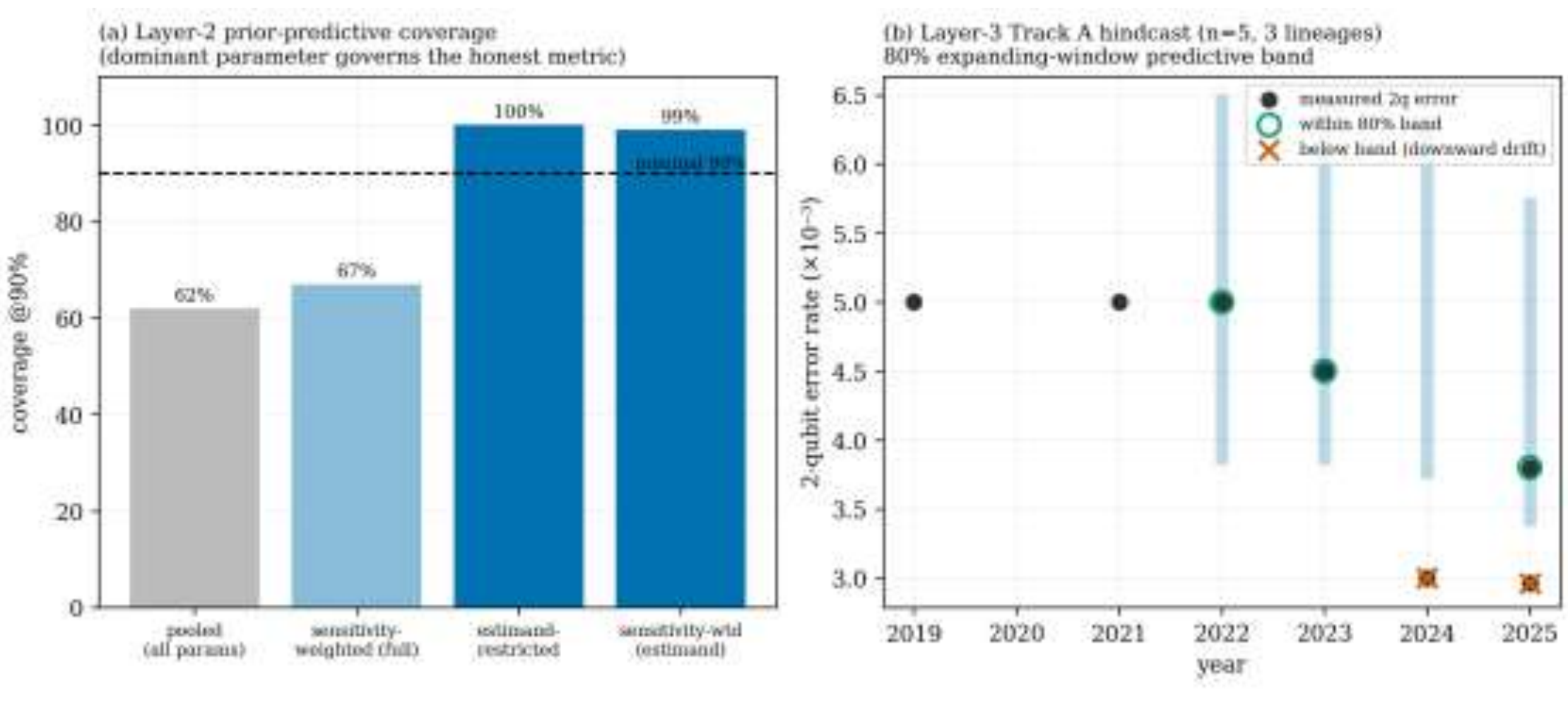

*Figure 4. (a) Layer-2 prior-predictive coverage at nominal 90 percent for four metrics, where the sensitivity-weighted metric, governed by the dominant parameter, reaches about 99 percent. (b) Layer-3 Track A expanding-window hindcast over five cases and three device lineages, with the 80 percent predictive band. The two 2024 to 2025 misses are downward*

*technological drift. The 2019 Sycamore point is on the average-gate estimand (simultaneous Pauli error times four fifths), and the 2023 IBM Heron point is a manufacturer-reported approximate value.*

Layer 1 confirms that the numerical machinery is correct. Across four analytic benchmarks with closed-form output distributions and all three samplers, all twelve cases pass, with every 5, 50, and 95 percent quantile within the pre-registered two percent tolerance and every mean within one percent. Any downstream effect is therefore a property of the priors and models, not a sampling bug.

Layer 2 tests prior-predictive coverage by leave-one-source-out. The honest headline is the sensitivity-weighted metric, which weights each parameter's coverage by how much it drives the output interval. This reaches about 99 percent at nominal 90 percent when restricted to the dominant error-rate estimand, and the error-rate estimand itself is covered four times out of four. The pooled coverage across all parameters is lower, at 62 percent, but this figure weights the dominant error rate the same as parameters with almost no output effect, such as the decoder reaction time, and its shortfall comes entirely from those low-sensitivity parameters. The calibration that the paper actually claims, namely the calibration of the resource interval, is governed by the error rate, and the error rate is well covered within its estimand.

Layer 3 is exploratory by design, because no fault-tolerant machine has executed these workloads. The Track A hindcast fits an error-rate prior from measurements before a cutoff year and asks whether the values reported at the cutoff fall inside the predictive interval. With the dated time series this yields five cases across three device lineages, namely Google, IBM, and USTC. Three of the five fall within the 80 percent band, with a Wilson interval of 27 to 86 percent that contains the nominal level, and three fall within the 90 percent band. The two misses, both from 2024 to 2025, are downward. Error rates fell faster than a static prior fitted from earlier years would predict. Under the pre-registered decomposition these are classified as technological-improvement drift rather than miscalibration, since a static epistemic prior does not forecast the pace of hardware progress. The Track B check looks at documented RSA-2048 physical-qubit estimates across generations, from about $10^9$ in 2012 to about $2 \times 10^7$ in 2019 to 2021 and below $10^6$ in 2025. That reduction is dominated by algorithmic and architectural innovation, which is exactly the axis the framework does not model, so a

## 5. Discussion

The three quantitative findings point to a single practical conclusion. A fault-tolerant resource estimate should be reported as an interval, accompanied by the fraction of the evidence-based parameter space that the estimating tool cannot represent, and by the error-rate assumption it was computed under. A single number quoted at the $10^{-3}$ planning convention conveys none of this. For someone timing the migration to post-quantum cryptography, the difference is not academic. At measured superconducting error rates the median physical-qubit requirement for RSA-2048 is about 4.6 times the conventional point value, and the 90% interval spans about forty times. The width is not incidental. It arises because the required code distance grows steeply as the physical error rate approaches the surface-code threshold, so the same prior spread in the error rate maps to a far wider spread in physical qubits. A migration deadline derived from the optimistic point estimate is therefore anchored to the low end of a wide and upward-shifted distribution. The same caution applies to claims of quantum advantage, where the roughly twofold disagreement among cost models means that any single-tool estimate carries a model-form uncertainty band that a point value hides.

The censoring result deserves a careful reading. It is not a claim that these workloads are physically impossible, and it is not a defect of the two tools we used. Both estimators are engineered around the planning convention. The Azure estimator caps the code distance at a value that is comfortable at $10^{-3}$. The Qualtran implementation of the Gidney and Fowler model uses a fixed distillation factory whose error budget is adequate at $10^{-3}$ but exceeded once the physical error rate rises above roughly $1.5 \times 10^{-3}$. Two independently built tools, provisioned by their authors for the conventional operating point, both refuse most of the evidence-based parameter space. The useful observation is the mismatch between what hardware has demonstrated and what standard planning tools are configured to represent. The tools can be reconfigured, and a practitioner who raises the distance cap or enlarges the factory will move the boundary. The point is that their default outputs, computed at convention values, are not summaries of the measured evidence.

Two features make the framework useful beyond the specific numbers reported here. It is tool-agnostic, so incorporating a new estimator is a matter of writing one adapter to the shared schema rather than re-encoding the uncertainty analysis. And the prior library is versioned, so it can be refitted as the hardware record grows. The Layer-3 hindcast already shows the direction of travel. Measured error rates have fallen faster than a static prior would predict, which means that as the evidence base deepens the priors will tighten and the resulting intervals will narrow. The contribution is therefore a reusable instrument that produces a living estimate, not a single verdict fixed at today's evidence.

Finally, we are precise about what the word evaluated means in this paper, because no fault-tolerant machine has executed any of these workloads and realised ground truth is unobtainable by anyone. The confirmatory weight rests on the first two evaluation layers. Layer 1 shows the numerical propagation is correct. Layer 2 shows that the calibration the paper actually claims, the calibration of the resource interval, is governed by the physical error rate. That dominant parameter is well covered within its estimand, with a sensitivity-weighted coverage near the nominal level. Layer 3 is exploratory, and its main value is to delimit scope rather than to confirm the intervals. Its cross-generation check demonstrates that the large historical reductions in resource estimates were driven by algorithmic and architectural innovation, which the framework deliberately does not model, so a parametric interval neither would nor should have predicted them. We therefore claim that the intervals are evaluated in a specific and limited sense, that the dominant parameter's calibration is tested and holds, and not that the full resource distribution has been validated against reality.

## 6. Limitations

We state the limitations explicitly.

1. **Single platform and architecture.** All results are for the superconducting two-dimensional nearest-neighbour surface code. The prior library carries neutral-atom and trapped-ion entries, and the framework is platform-tagged. Those platforms are not exercised here, because they do not yet have enough independent large-array error-rate measurements to support an evidence-based prior for the dominant parameter. Extending the study is future work gated on evidence, not on new machinery.

2. **Retrospective and thin Layer 3.** The temporal hindcast has five cases across three device lineages and is labelled exploratory in the pre-registration. It cannot carry confirmatory weight, and we do not ask it to. The confirmatory claims rest on Layers 1 and 2.

3. **Design-dependent priors are wide by construction.** The routing and factory-throughput priors span a broad range because they encode genuine design freedom rather than a measured quantity. Their low leave-one-source-out coverage is the intended behaviour for a wide, design-space prior, not a calibration failure, and because these parameters carry almost no output sensitivity, their width does not affect the resource interval.

4. **The survivorship factor is a judgement.** The factor of 1.3 that shifts the error-rate prior worse than the published record is a modelling choice rather than a measurement. We do not fix it, and we report the sensitivity of the conclusions to it. The headline effects survive its removal. Set the factor to one, so that the prior trusts the published record exactly, and the error-rate median is still about $3 \times 10^{-3}$, close to three times the $10^{-3}$ convention. The median RSA-2048 requirement then sits about 2.9 times above the convention point rather than 4.6 times, and the 90% interval still spans about 23 times rather than forty. Dropping the factor shrinks the magnitudes. It does not touch the two conclusions, the upward shift and the wide interval.

5. **The structural spread is a lower bound.** The five models share the same code-distance physics and differ only in layout and factory accounting, so the factor-of-two spread measures model-form disagreement about those choices. The true structural uncertainty could be larger if the shared physics itself is uncertain.

6. **Two inputs are provisional.** The Hubbard-model workload is an order-of-magnitude proxy rather than an exact construction, and the 2023 IBM Heron point in the Layer-3 time series is a manufacturer-reported approximate value. Neither affects the four headline findings.

## 7. Conclusion

Propagating the error rates that superconducting hardware has actually demonstrated, rather than the optimistic planning convention, makes the canonical fault-tolerant workloads both markedly more expensive and far less certain than a point estimate implies. The 90% resource interval for RSA-2048 spans about forty times, and its median sits several times above the conventional value. Five independently structured cost models disagree by a stable factor of two, with two independent third-party tools falling inside that band. Both of those tools exceed their representable envelope for the large majority of the evidence-based parameter space. We provide the framework that produces these results as an open, tool-agnostic, and pre-registered instrument, together with the documented prior library and the analysis code. As hardware improves and the evidence base deepens, the same instrument can be refitted, so its intervals are intended to be a living estimate that tightens with the field rather than a fixed claim. For any decision that depends on when a fault-tolerant machine will be able to run a given algorithm, the honest input is an evaluated interval with its representable fraction stated, not a single optimistic number.

## Author contributions

Furqan Nasir conceived the study, designed the framework and the pre-registered evaluation, implemented the software and the numerical experiments, and drafted the manuscript. Sher Jeel Ahmad contributed to the interpretation of the fault-tolerant cost models and reviewed the manuscript. All authors approved the final version.

## Use of AI tools

The authors conceived and directed this study, made the scientific and methodological decisions, verified every hardware parameter against its primary source, and take full responsibility for the entire manuscript, including all text, figures, code, and claims. In carrying out the work they used a generative AI assistant (a large language model) to help implement and debug the analysis software, run and organise the numerical experiments, produce figures and tables from the computed results, search the literature, and draft and language-edit the text. All quantitative results were produced by the openly released code and were checked by the authors, and no AI-generated content was included without author review and verification. The assistant is not an author and bears no responsibility for the work.

## Funding

This work received no external funding.

## Competing interests

The authors declare no competing interests.

## Data and code availability

The framework (`qre-uq`), the evidence-based prior library, the self-contained analysis notebook, and the figure-generation scripts are openly available. The evaluation protocol was pre-registered at the Open Science Framework (osf.io/3d9m2, registered 19 July 2026) before the confirmatory analyses were run. An archived snapshot of the code and prior library is deposited at Zenodo (DOI: 10.5281/zenodo.21463123). All figures and tables in this paper can be regenerated from the released data with the accompanying scripts, using the fixed random seeds recorded in Appendix D.

## Acknowledgements

The authors thank the developers of the open-source estimators and libraries used as third-party tools in this study, in particular the Azure Quantum Resource Estimator and Qualtran, and the hardware groups whose published characterisations make an evidence-based prior possible.

## Appendix A. Prior library and evidence base

The prior for each parameter is fitted from a documented evidence table (`evidence.csv`) by weighted maximum likelihood, with dependency grouping to avoid double-counting non-independent sources. The number of independent evidence groups per parameter, on the superconducting platform, is: physical error rate, six; routing overhead, three; decoder reaction time, three; cycle time, two; factory throughput, two. The dominant parameter, the physical error rate, is defined on the estimand of the average two-qubit gate error of a scalable two-dimensional surface-code array. Its six sources are the five large-array devices listed in Section 3 plus a down-weighted small-scale fluxonium record and the optimistic planning convention. All sources are standardised to the average-gate estimand before fitting. Where a source reports a Pauli error, it is converted using the two-qubit factor of four fifths. The design-dependent parameters, routing overhead and factory throughput, are given wide priors by construction because they encode design freedom rather than a measured quantity.

## Appendix B. Judgement-call log

The study records its non-obvious modelling choices so that they can be inspected and varied.

- **Error-rate estimand.** The average two-qubit gate error, not the Pauli error or a device-record best value. Pauli errors are converted by the four-fifths factor.
- **Decoder parameter.** The decoder reaction time is treated as a software latency per round, not a throughput. It carries near-zero output sensitivity.
- **Survivorship factor.** A factor of 1.3 shifts the error-rate prior worse than the published record. It is swept rather than fixed.
- **AES-256 accounting.** AES-256 is quoted in T gates (the full-Grover figure of 1.36 × 2^145). Tools that consume T states use the standard four-T-per-Toffoli conversion in their factory accounting.
- **Azure QRE envelope.** The code distance is capped at fifty by default. Draws that require a larger distance are recorded as censored rather than dropped.
- **Qualtran configuration.** The Gidney and Fowler CCZ2T model uses a fixed distillation factory with level-one and level-two distances of fifteen and thirty-one, and selects the data-block distance from the residual error budget. The routing overhead is mapped from our multiplicative convention to Qualtran's additive convention by subtracting one.

- **Hubbard workload.** Included as an order-of-magnitude simulation proxy, and labelled as such.

## Appendix C. Pre-registration and deviations

The evaluation was registered at the Open Science Framework (osf.io/3d9m2) on 19 July 2026, before the confirmatory analyses were run, and is under embargo until 2028. The registration fixes the three-layer design, the estimand definition, the source-weighting rules, and the labelling of Layers 1 and 2 as confirmatory and Layer 3 as exploratory. We record the following deviations and additions for full transparency.

1. The 2019 Sycamore point in the Layer-3 time series was standardised to the average-gate estimand (simultaneous Pauli error times four fifths) after registration, to keep the whole series on one estimand. This is a consistency correction, documented in the claims ledger.
2. The deepened Track-A time series is an addition beyond the minimal pre-registered rule of testing only cutoffs with at least two prior sources. It is reported alongside, not in place of, the registered analysis, and is labelled exploratory.
3. The calibration and censoring-robustness analyses are post-registration supporting analyses. They are descriptive robustness checks, not confirmatory tests, and are labelled as such.

## Appendix D. Reproducibility

All results are reproducible from the released code with fixed seeds. Parameter draws use inverse- transform sampling with Latin-hypercube or Sobol sequences and a fixed seed of zero. The parametric intervals use two million analytic draws. The third-party-tool comparison uses eighty thousand shared draws per workload, identical across tools. The third-party estimators are the Azure Quantum Resource Estimator, run locally through its Q# package, and Qualtran's Gidney and Fowler CCZ2T model, accessed through the verified `get_ccz2t_costs_from_error_budget` interface. The full third-party run is executed in a single free-tier CPU session (four cores) in the self-contained notebook `notebooks/qre_uq_qualtran_kaggle.ipynb`, with numpy and scipy pinned during installation to preserve the numerical stack. Every figure and table is regenerated by `paper/figures/make_figs.py`, which reads the run outputs in `results/kaggle_qualtran/` and re-derives the transparent-model distributions from the frozen priors. The script checks, before drawing, that the regenerated RSA-2048 analytic median reproduces the frozen two-million-draw value.

## Appendix E. Supplementary tables

**Table E1. Analytic-model to Azure QRE ratios, by error-rate bin (uncensored region, RSA-2048).** The transparent models track the calibrated tool by a stable factor across the error-rate range.

| Error-rate bin | n | analytic / AQRE | Litinski / AQRE | Fowler / AQRE |
|---|---|---|---|---|
| $[1.0\times10^{-4}, 1.5\times10^{-3})$ | 59 | 0.75 | 1.50 | 1.21 |
| $[1.5\times10^{-3}, 2.5\times10^{-3})$ | 383 | 0.74 | 1.49 | 1.17 |
| $[2.5\times10^{-3}, 9.0\times10^{-3})$ | 35 | 0.73 | 1.46 | 1.19 |
| all uncensored | 477 | 0.74 | 1.48 | 1.17 |

**Table E2. Azure QRE censored fraction versus error budget (RSA-2048).** Loosening the budget fifty times lowers censoring by only ten points.

| Error budget | 0.01 | 0.05 | 0.10 | 0.20 | 0.33 | 0.50 |
|---|---|---|---|---|---|---|
| % censored | 81 | 77 | 75 | 73 | 72 | 71 |

**Table E3. Layer-3 Track A cases (expanding-window hindcast).** A tick means the value at the cutoff fell inside the prior-predictive interval fitted from earlier points. The two 2024–2025 misses are downward technological drift.

| Cutoff | Device | Value | 50% | 80% | 90% | Lineage |
|---|---|---|---|---|---|---|
| 2022 | Google d=5 surface code | $5.0\times10^{-3}$ | ✓ | ✓ | ✓ | google |
| 2023 | IBM Heron r1 | $4.5\times10^{-3}$ | ✓ | ✓ | ✓ | ibm |
| 2024 | Google Willow | $3.0\times10^{-3}$ | — | — | — | google |
| 2025 | IBM Heron R2 | $2.96\times10^{-3}$ | — | — | — | ibm |
| 2025 | USTC Zuchongzhi-3 | $3.8\times10^{-3}$ | — | ✓ | ✓ | ustc |